# Spectroscopy of a κ-Cygnid fireball afterglow

José M. Madiedo[1, 2]

[1]Facultad de Ciencias Experimentales, Universidad de Huelva, Avda. de las Fuerzas Armadas S/N. 21071 Huelva, Spain.

[2]Depto. de Física Atómica, Molecular y Nuclear, Facultad de Física, Universidad de Sevilla, 41012 Sevilla, Spain.

Tel.: +34 959219991

Fax: +34 959219983

Email: madiedo@cica.es

**ABSTRACT**

A bright fireball with an absolute magnitude of -10.5 ± 0.5 was recorded over the South of Spain on August 15, 2012. The atmospheric trajectory, radiant and heliocentric orbit of this event are calculated. These data show that the parent meteoroid belonged to the κ-Cygnid meteoroid stream. The emission spectrum of this bolide, which was obtained in the wavelength range between 350 and 800 nm, suggests a chondritic nature for the progenitor meteoroid. Besides, the spectrum of the meteoric afterglow was also recorded for about 0.7 seconds. The evolution with time of the intensity of the main emission lines identified in this signal is discussed.

**1 INTRODUCTION**

Bright fireballs, specially those moving at high velocity, may produce long-lasting glows called persistent trains. These phenomena, which often form as a consequence of bright meteor flares, can be visible for several minutes after the meteor has disappeared. Once it is formed, the luminosity of the persistent train falls quickly within a few seconds during the so-called afterglow phase.

Meteor spectroscopy is a fundamental technique to get data about the physicochemical composition in meteoric plasmas, and also to get an insight about the chemical composition of meteoroids ablating in the atmosphere (Borovička 2003, Jenniskens 2007). Besides, the analysis of fireball afterglow spectra can provide useful information about the physical processes taking place in persistent meteor trains. However these afterglow spectra are not abundant in the literature (see e.g. Borovička & Jenniskens 2002, Abe et al. 2004, Jenniskens et al. 2000, Madiedo et al. 2014a).

Despite the κ-Cygnids do not move at very high speeds, since these meteoroids impact the atmosphere with a velocity of about 25 km s$^{-1}$, bright κ-Cygnids meteors tend to exhibit a final flare as a consequence of the sudden disruption of the progenitor meteoroid when the particle enters denser atmospheric regions (see e.g. Trigo-Rodríguez et al. 2009). On 15 August 2012 at $23^h44^m59.7^s$ UTC, a κ-Cygnid fireball with an absolute magnitude of -10.5 ± 0.5 was simultaneously recorded from two meteor observing stations located in the South of Spain. The emission spectrum of the bolide was also recorded. The event reached its maximum luminosity during a very bright flare that took place by the end of its atmospheric path, giving rise to a persistent train. The emission spectrum of the meteoric afterglow was recorded during about 0.7 seconds. This paper focuses on the analysis of this spectrum afterglow. The emission spectrum of the fireball is also discussed. Besides, the atmospheric trajectory and radiant of this fireball are calculated, and the orbital parameters and tensile strength of the progenitor meteoroid are obtained.

## 2 INSTRUMENTATION AND METHODS

The fireball discussed in this paper was recorded from two meteor observing stations located in the South of Spain: Sevilla (latitude: 37º 20' 46" N, longitude: 5º 58' 50" W, height: 28 m), and El Arenosillo (latitude: 37º 06' 16" N, longitude: 6º 43' 58" W, height: 40 m). These stations employ an array of low-lux monochrome CCD cameras (models 902H2 and 902H Ultimate, manufactured by Watec Co.) that generate interlaced video imagery at 25 frames per second (fps) with a resolution of 720x576 pixels. Full details about the operation of this array of video cameras are given in Madiedo & Trigo-Rodríguez (2008) and Madiedo et al. (2010). For data reduction the AMALTHEA software was employed (Madiedo et al. 2011, 2013a), which calculated the fireball atmospheric trajectory, radiant position and meteoroid orbital data by following the methods described in Ceplecha (1987).

To obtain meteor spectra, holographic diffraction gratings (with 1000 grooves/mm) are attached to the objective lens of some of the CCD video cameras that operate at the above-mentioned stations. Their spectral response is shown in Figure 1. These slitless videospectrographs operate in the framework of the SMART Project, which was started in 2006 (Madiedo 2014). The spectra have been analyzed with the CHIMET software (Madiedo et al. 2013b, 2014b; Madiedo 2014, 2015).

**3 OBSERVATIONS AND RESULTS**

**3.1 Atmospheric path and meteoroid orbit**

The atmospheric trajectory of the fireball was triangulated from the analysis of the video images obtained by the CCD cameras operating at Sevilla and El Arenosillo. According to this, the meteoroid impacted the atmosphere with a velocity $V_\infty = 27.3 \pm 0.3$ km s$^{-1}$ and with an inclination of 26.4º with respect to the local vertical. The fireball began at a height of $109.7 \pm 0.5$ km above the sea level, and ended at a height of $72.0 \pm 0.7$ km. When the fireball was located at a height of $75.2 \pm 0.7$ km, it exhibited a bright flare. At this stage the event reached its maximum luminosity, which corresponded to an absolute magnitude of $-10.5 \pm 0.5$. Table 1 shows the main parameters of this atmospheric path and the position of the geocentric radiant (J2000). The orbital parameters of the progenitor meteoroid are listed in Table 2. These data confirm the association of this event with the κ-Cygnid meteoroid stream. Thus, the value of the Southworth & Hawkins $D_{SH}$ dissimilarity function (Southworth & Hawkins 1963) obtained by comparing the orbit of this stream (Sekanina 1973) with the orbit of the meteoroid yields $D_{SH} = 0.08$. This value remains below the cut-off value of 0.15 usually adopted to establish a valid association (Lindblad 1971a, 1971b).

**3.2 Fireball spectrum.**

The fireball spectrum is shown in Figure 2. It was obtained in the wavelength range between 350 and 800 nm with a video spectrograph operating at station #1. The signal, which was initially obtained on each video frame as an intensity profile (in arbitrary units) versus pixel number, was converted into intensity versus wavelength by indentifying some of the emission lines in the spectrum. This calibration in wavelength was performed by using the contributions from Na I-1 (588.9 nm) and Mg I-2 (517.2 nm), where Moore's multiplet numbers have been employed (Moore 1945). Then, the contributions from different frames were added to get an integrated spectrum along the whole meteor path. This spectrum was corrected by taking into consideration the spectral sensitivity of the spectrograph shown in Figure 1. Only frames corresponding to the bright flare exhibited by the fireball by the end of its atmospheric path were excluded from this analysis, since most of the signal was saturated at that stage and so it was of very limited use. In Figure 2, the main contributions identified in the fireball spectrum have been indicated. As usual in meteor spectra, most of the lines identified in the signal correspond to Fe I. The strongest emissions caused by the meteoroid in this spectrum are those from Fe I-5 at 374.5 nm, the Na I-1 doublet at 588.9 nm and the Mg I-2 triplet at 517.2 nm. Several contributions from atmospheric nitrogen and oxygen are also present. Thus, $N_2$ bands have been identified in the red region of the spectrum, although the most important atmospheric contribution corresponds to the O I line at 777.4 nm.

### 3.3 Afterglow spectrum.

The same videospectrograph that recorded the fireball spectrum also recorded the emission spectrum of the persistent train produced after the bright flare that took place next to the end of the atmospheric path of the bolide. This spectrum corresponded to the afterglow phase, and it was recorded for around 0.7 s after the formation of the persistent train. The total duration of the train is unknown, since the cameras that recorded the event are configured to stop recording 1 second after the fireball ends. The brightest emission lines in this signal were saturated at the beginning of the train formation. So, the actual intensity of these lines could only be measured after some time. Figure 3 shows the afterglow spectrum at t = 0.34 s after the formation of the persistent train. The afterglow spectrum covered the range between 400 and 800 nm, since the region in the ultraviolet between 350 and 400, which could be recorded for the fireball spectrum, was out of the field of view of the camera at the moment of the train formation. This signal was calibrated in wavelength and corrected for the spectral sensitivity of the recording device by following the same technique employed for the fireball spectrum.

The most important contributions in the afterglow spectrum correspond to multiplets Na I-1 (589 nm), Fe I-1 (511-517 nm), Fe I-2 (437-449 nm) and Ca I-2 (422.6 nm). The emissions due to Fe I-15 (527-545 nm) and Na I-6 (568 nm) were also identified, together with those of Ca I-3 (610-616 nm) and Ca I-1 (657 nm). The contribution of FeO "orange arc" emission between 550 and 650 nm is also very likely, as was found by Jenniskens et al. (2000) and Abe et al. (2005) in the afterglow spectrum of Leonid fireballs.

## 4 DISCUSSION
### 4.1 Meteoroid initial mass and strength

The initial (pre-atmospheric) mass of the meteoroid was estimated from the lightcurve of the fireball, which is shown in Figure 4, by using the classical meteor luminous equation:

$$m_\infty = 2 \int_{t_b}^{t_e} I/(\tau v^2) \, dt \tag{1}$$

where $m_\infty$ is the above-mentioned mass, $t_b$ and $t_e$ the time instants corresponding to the beginning and the end of the luminous trajectory of the meteor, respectively. And I is the time-dependent measured luminosity of the event, which is given as a function of its absolute magnitude M by means of the equation

$$I = 10^{-0.4 \cdot M} \tag{2}$$

The velocity-dependent luminous efficiency $\tau$ (i.e., the fraction of the kinetic energy of the meteoroid that is converted into light) was calculated from the relationships given by Ceplecha & McCrosky (1976). In this way, the mass of the meteoroid yields 330 ± 50 g. Its diameter, by considering that the bulk density of κ-Cygnid meteoroids is of about 2.2 g cm$^{-3}$ (Babadzhanov & Kokhirova 2009), yields 6.6 ± 0.3 cm.

The bright flare exhibited by the fireball by the end of its atmospheric path took place as a consequence of the sudden disruption of the meteoroid. This flare has been employed to obtain the tensile strength of this particle by following the procedure described in Trigo-Rodríguez & Llorca (2006). According to this approach, meteoroid strengths can be estimated as the aerodynamic pressure P at the disruption point (Bronshten 1981):

$$P = \rho_{atm} \cdot v^2 \qquad (3)$$

where v and $\rho_{atm}$ are the velocity of the meteoroid and the atmospheric density at the height where this fracture takes place, respectively. Here the atmospheric density was calculated by following the US standard atmosphere model (U.S. Standard Atmosphere 1976). Since the flare took place at a height of 75.2 ± 0.7 km above the sea level and the velocity of the meteoroid, obtained from the analysis of its atmospheric path, was 26.7 ± 0.9 km s$^{-1}$ at that position, the calculation of the strength yields 24 ± 5 kPa. This value is very similar to the strength found for stronger κ-Cygnid meteoroids involved in the outburst experienced by this stream in 2007 by Trigo-Rodríguez et al. (2009) (18 ± 2 kPa).

### 4.2 Fireball spectrum

To get an insight into the chemical nature of the meteoroid, the approach employed in (Borovička et al. 2005) was employed. So, the relative intensity of the Na I-1, Mg I-2 and Fe I-15 multiplets were measured in the fireball spectrum, and the value of the Na/Mg and Fe/Mg intensity ratios were calculated. These ratios yield Na/Mg = 0.99 ± 0.05 and Fe/Mg = 1.13 ± 0.09. This Na/Mg intensity ratio is in good agreement with the expected result (Na/Mg ≈ 1) for meteoroids with chondritic composition when the meteor velocity is of about 27 km s$^{-1}$ (Borovička et al. 2005). The relative intensities of these three multiplets are plotted on the ternary diagram shown in Figure 5, where the solid curve shows the expected relative intensity as a function of meteor velocity for chondritic meteoroids (Borovička et al. (2005)). The position on this plot of the point describing the fireball spectrum shows that the progenitor meteoroid can be regarded as normal, according to the classification given by Borovička et al. (2005). As can be noticed, this experimental value fits fairly well the expected relative intensity for chondritic meteoroids for a meteor velocity of about 27 km s$^{-1}$.

### 4.3 Afterglow spectrum

Two known mechanisms have been identified as responsible for the emission of radiation in fireball afterglows: cooling of hot rarified gas and recombination (Borovička 2006). In some fireballs both mechanisms are present and form two subsequent phases of train radiation, while in other trains only one of them can be observed. The spectra of both phases do not differ dramatically, the main difference being the presence of the Mg I-2 line in the recombination phase (Borovička 2006). The fact that this line was not identified in the afterglow spectrum of the fireball analyzed here, points to the cooling mechanism as responsible for the observed afterglow spectrum. The relative intensity of the main emission lines in the calibrated afterglow spectrum (those of multiplets Na I-1 (588.9 nm), Fe I-1 (516.9 nm), Ca I-2 (422.6 nm) and Fe I-2 (448.2 nm)) were measured. Their dependence with time is shown in Figure 6. This plot suggests an exponential decay of the relative intensity I of these lines:

$$I = I_0 \exp(B \cdot t) \tag{4}$$

This exponential decay was also found for the afterglow spectrum of Leonid fireballs (Borovička & Jenniskens 2000), but also for sporadic bolides (Madiedo et al. 2014a). By fitting the intensity of each multiplet in Figure 6 to Eq. (4), the corresponding values of the parameters $I_0$ and B have been calculated. These are listed in Table 3, which clearly shows that the parameter B, which measures the decay rate, decreases as the excitation potential $E_k$ increases. This behaviour suggests that the observed decrease in luminosity in these emission lines in the afterglow spectrum is mainly controlled by a temperature-driven mechanism in the meteor train (Borovička & Jenniskens 2000). The fact that the decay is faster for higher values of $E_k$ is also shown in Figure 7, which suggests that this dependence can be described, within the experimental uncertainty, by means of a linear equation:

$$B = B_0 + DE_k \tag{5}$$

where D is the so-called cooling constant. By fitting the values of B in Table 3 to Eq. (5), one obtains $B_0 = 0.3$ $s^{-1}$ and $D = -1.6$ $s^{-1}eV^{-1}$. The value of D was found to be of -1.5 $s^{-1}eV^{-1}$ for a mag. -13 Leonid fireball (Borovička & Jenniskens 2000), and -2.7 $s^{-1}eV^{-1}$ for two additional mag. -8 Leonid bolides (Abe et al. 2005). So, the cooling was slightly faster in the persistent train of the κ-Cygnid analyzed here that in the persistent train of the above-mentioned mag. -13 Leonid fireball, but slower that for the mag. -8 Leonid bolides.

# 5 CONCLUSIONS

The analysis of the double-station fireball discussed in this work has provided the atmospheric trajectory and radiant of the bolide, and also the heliocentric orbit of the progenitor meteoroid. These data reveal that this particle, which had a pre-atmospheric mass of 330 ± 50 g, belonged to the κ-Cygnid meteoroid stream. The luminous phase of the bolide started at 109.7 ± 0.5 km above the sea level and had its terminal point at a height of 72.0 ± 0.7 km. The fireball reached its maximum brightness during a bright flare that took place by the end of its atmospheric path, reaching an absolute magnitude of -10.5 ± 0.5. The tensile strength of the meteoroid, which was estimated from the air density and meteoroid velocity measured at this flare, yields 24 ± 5 kPa, which is similar to the strength previously found for stronger members of the κ-Cygnid meteoroid stream.

In the fireball spectrum, emission lines produced by Na I, Fe I, Mg I and Ca I were identified, together with different emissions from atmospheric oxygen and nitrogen. This spectrum suggests a chondritic nature for the meteoroid. On the other hand, several emission lines produced by Na, Ca and Fe were identified in the afterglow emission spectrum, which could be recorded for about 0.7 seconds. The brightness of the emission lines of Na I-1, Fe I-1, Ca I-2 and Fe I-2 was found to decrease exponentially with time. The observations suggest that this decrease in luminosity is mainly controlled by a temperature-driven mechanism in the meteor train.

## ACKNOWLEDGEMENTS
I thank Dr. Jiri Borovička and an anonymous referee for their valuable comments and suggestions.

## REFERENCES
Abe S., Ebizuka N., Murayama H., Ohtsuka K., Sugimoto S. et al., 2004. Video and Photographic Spectroscopy of 1998 and 2001 Leonid Persistent Trains from 300 to 930 nm. Earth Moon Planets, 95, 265-277. doi: 10.1007/s11038-005-9031-0

Babadzhanov P.B., Kokhirova G.I., 2009. Densities and porosities of meteoroids . A&A, 495, 353-358. doi: 10.1051/0004-6361:200810460

Borovička J., 1993. A fireball spectrum analysis. A&A, 279, 627-645.

Borovicka J., 2006, Meteor Trains - Terminology and Physical Interpretation. Journal of the Royal Astronomical Society of Canada, 100, 194-198.

Borovička J. and Jenniskens P., 2000. Time Resolved Spectroscopy of a Leonid Fireball Afterglow. Earth Moon Planets, 82/83, 399-428.

Borovička J., Koten P. Spurny P., Boček J. and Stork R., 2005. A survey of meteor spectra and orbits: evidence for three populations of Na-free meteoroids. Icarus, 174, 15-30. doi: 10.1016/j.icarus.2004.09.011

Bronshten V.A., 1981, Geophysics and Astrophysics Monographs. Reidel, Dordrecht.

Ceplecha Z., 1987. Geometric, dynamic, orbital and photometric data on meteoroids from photographic fireball networks. Bull. Astron. Inst. Cz., 38, 222-234.

Ceplecha Z. & McCrosky R.E., 1976. Fireball end heights - A diagnostic for the structure of meteoric material. Journal of Geophysical Research, 81, 6257-6275.

Jenniskens P., Lacey M., Allan B.J., Self D.E., Plane J.M.C., 2000. FeO "Orange Arc" Emission Detected in Optical Spectrum of Leonid Persistent Train. Earth Moon Planets, 82-83, 429-438.

Jenniskens P., 2007. Quantitative meteor spectroscopy: Elemental abundances. Advances in Space Research, 39, 491-512. doi: 10.1016/j.asr.2007.03.040

Lindblad B.A., 1971a. A stream search among 865 precise photographic meteor orbits. Smiths. Contr. Astrophys., 12, 1-13.

Lindblad B.A., 1971b. A computerized stream search among 2401 photographic meteor orbits. Smiths. Contr. Astrophys., 12, 14-24.

Madiedo J.M., Trigo-Rodriguez J.M., 2008. Multi-station video orbits of minor meteor showers. Earth Moon Planets, 102, 133-139. doi: 10.1007/s11038-007-9215-x

Madiedo J.M., Trigo-Rodríguez J.M., Ortiz J.L., Morales N., 2010. Robotic Systems for Meteor Observing and Moon Impact Flashes Detection in Spain. Advances in Astronomy, 2010, 1, doi: 10.1155/2010/167494

Madiedo J.M., Trigo-Rodríguez J.M., Lyytinen E. Data Reduction and Control Software for Meteor Observing Stations Based on CCD Video Systems. NASA/CP-2011-216469, 330, 2011.

Madiedo J.M., Trigo-Rodríguez J.M., Konovalova N., Williams I.P., Castro-Tirado A.J., Ortiz J.L., Cabrera J, 2013a. The 2011 October Draconids outburst - II. Meteoroid chemical abundances from fireball spectroscopy. MNRAS, 433, 571-580. doi: 10.1093/mnras/stt748

Madiedo J.M., Trigo-Rodríguez J.M., Lyytinen E., Dergham J., Pujols P., Ortiz J.L., Cabrera J., 2013b. On the activity of the γ-Ursae Minorids meteoroid stream in 2010 and 2011. MNRAS, 431, 1678-1685. doi: 10.1093/mnras/stt288

Madiedo J.M., 2014. Robotic systems for the determination of the composition of solar system materials by means of fireball spectroscopy. Earth, Planets and Space, 66, 70-79.

Madiedo J.M. et al., 2014a. Trajectory, orbit, and spectroscopic analysis of a bright fireball observed over Spain on April 13, 2013. A&A, 569, A104, doi:10.1051/0004-6361/201322120

Madiedo J.M., Trigo-Rodríguez J.M., Ortiz J.L., Castro-Tirado A.J., Cabrera-Caño, J., 2014b. Orbit and emission spectroscopy of α-Capricornid fireballs. Icarus, 239, 273-280.

Madiedo J.M., 2015. The ρ-Geminid meteoroid stream: orbits, spectroscopic data and implications for its parent body. MNRAS, 448, 2135-2140, doi: 10.1093/mnras/stv148

Moore C.E., 1945, In: A Multiplet Table of Astrophysical Interest. Princeton University Observatory, Princeton, NJ. Contribution No. 20.

Sekanina Z., 1973. Statistical Model of Meteor Streams. III. Stream Search Among 19303 Radio Meteors. Icarus, 18, 253-284.

Southworth R.B., Hawkins G.S., 1963. Statistics of meteor streams. Smithson Contr. Astrophys., 7, 261.

Trigo-Rodríguez J.M., Llorca J., 2006. The strength of cometary meteoroids: clues to the structure and evolution of comets. MNRAS, 372, 655-660. doi: 10.1111/j.1365-2966.2006.10843.x

Trigo-Rodríguez J.M., Madiedo J.M., Williams I.P., Castro-Tirado A.J., 2009. The outburst of the κ-Cygnids in 2007: clues about the catastrophic break up of a comet to produce an Earth-crossing meteoroid stream. MNRAS, 392, 367-375. doi: 10.1111/j.1365-2966.2008.14060.x

U.S. Standard Atmosphere, 1976, NOA-NASA-USAF, Washington.

TABLES

Table 1. Atmospheric trajectory and radiant data (J2000).

| $H_b$ (km) | $H_e$ (km) | $\alpha_g$ (°) | $\delta_g$ (°) | $V_\infty$ (km s$^{-1}$) | $V_g$ (km s$^{-1}$) | $V_h$ (km s$^{-1}$) |
|---|---|---|---|---|---|---|
| 109.7±0.5 | 72.0±0.7 | 291.5±0.3 | 60.6±0.2 | 27.3±0.3 | 25.0±0.3 | 37.7±0.3 |

Table 2. Orbital parameters (J2000).

| a (AU) | e | i (°) | q (AU) | ω (°) | Ω (°) |
|---|---|---|---|---|---|
| 2.70 ± 0.13 | 0.634 ± 0.017 | 41.0 ± 0.4 | 0.9895 ± 0.0007 | 199.7±0.3 | 143.35399±10$^{-5}$ |

Table 3. Calculated values of the parameters in Eq. (4) for the main emission lines identified in the afterglow spectrum. The excitation potential of the upper level ($E_k$) is included.

| Multiplet | λ (nm) | $E_k$ (eV) | $I_0$ (a.u.) | B (s$^{-1}$) |
|---|---|---|---|---|
| Na I-1 | 588.9 | 2.10 | 421 ± 4 | -2.9 ± 0.3 |
| Fe I-1 | 516.9 | 2.45 | 583 ± 7 | -3.8 ± 0.4 |
| Fe I-2 | 448.2 | 2.88 | 470 ± 8 | -4.1 ± 0.5 |
| Ca I-2 | 422.6 | 2.93 | 330 ± 4 | -4.4 ± 0.4 |

FIGURES

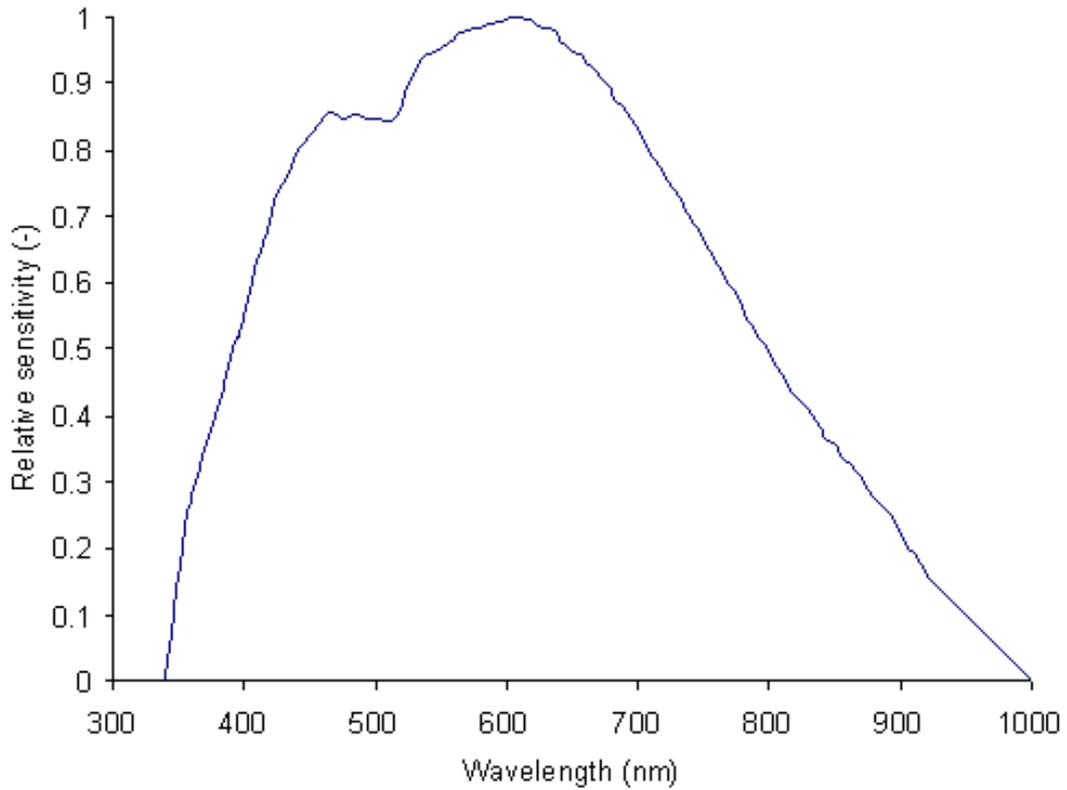

Figure 1. Relative spectral sensitivity of the spectrographs employed in this research.

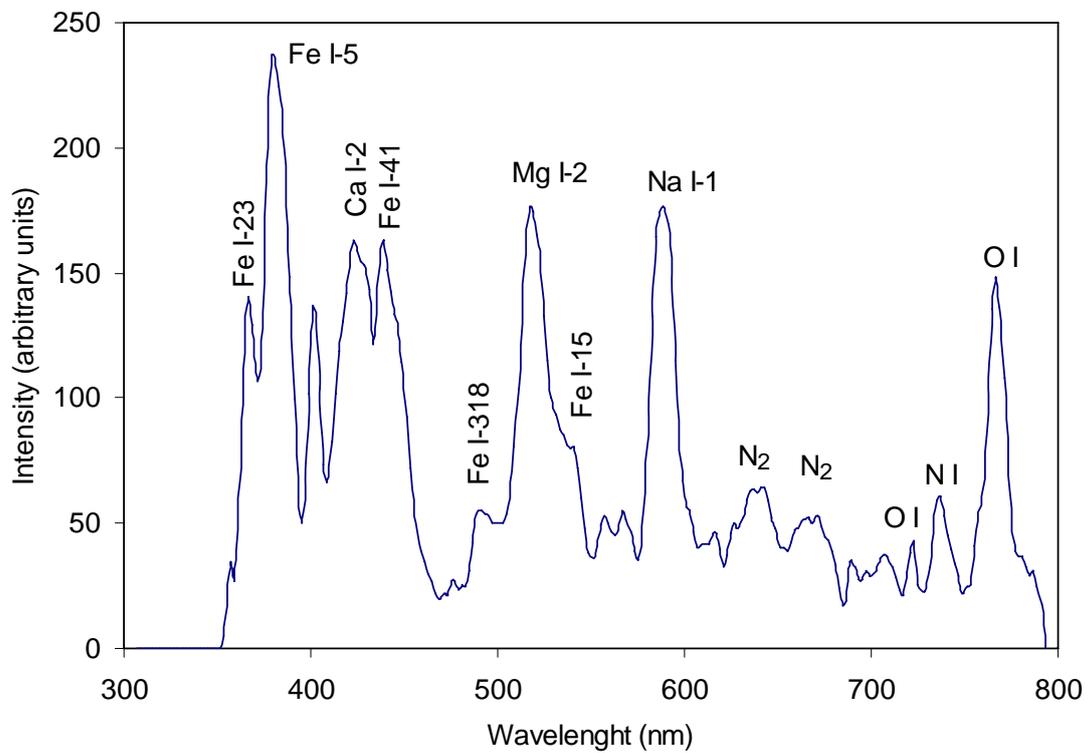

Figure 2. Calibrated emission spectrum produced by the fireball.

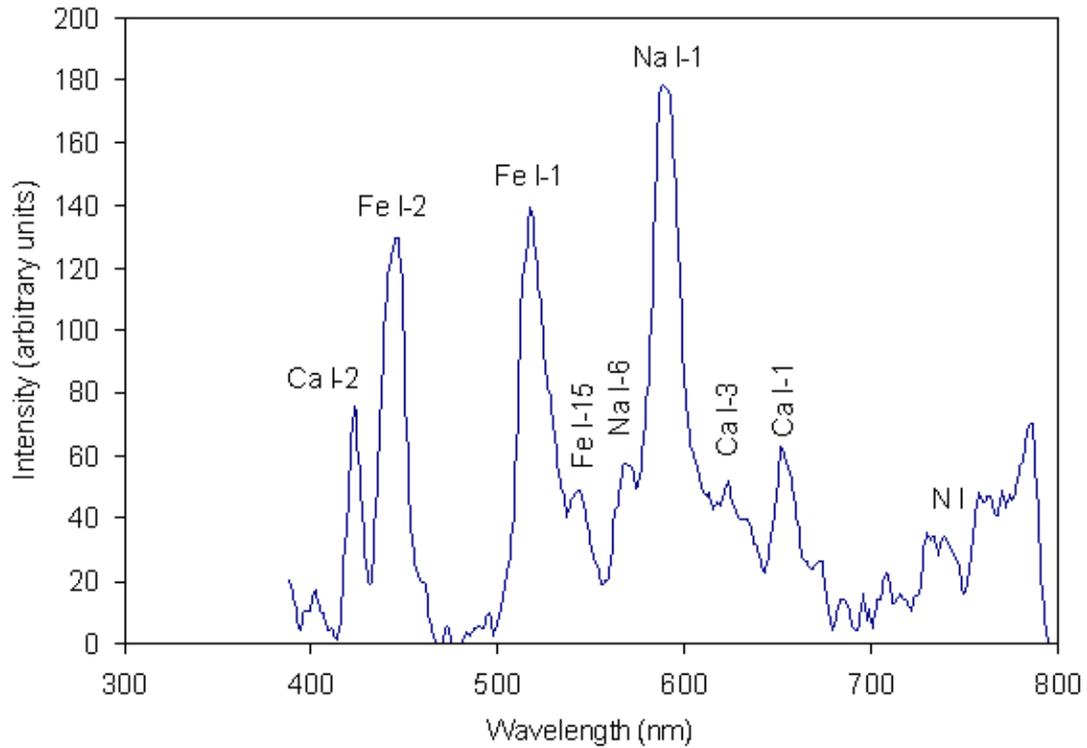

Figure 3. Calibrated afterglow spectrum at t=0.34 s after the formation of the persistent train.

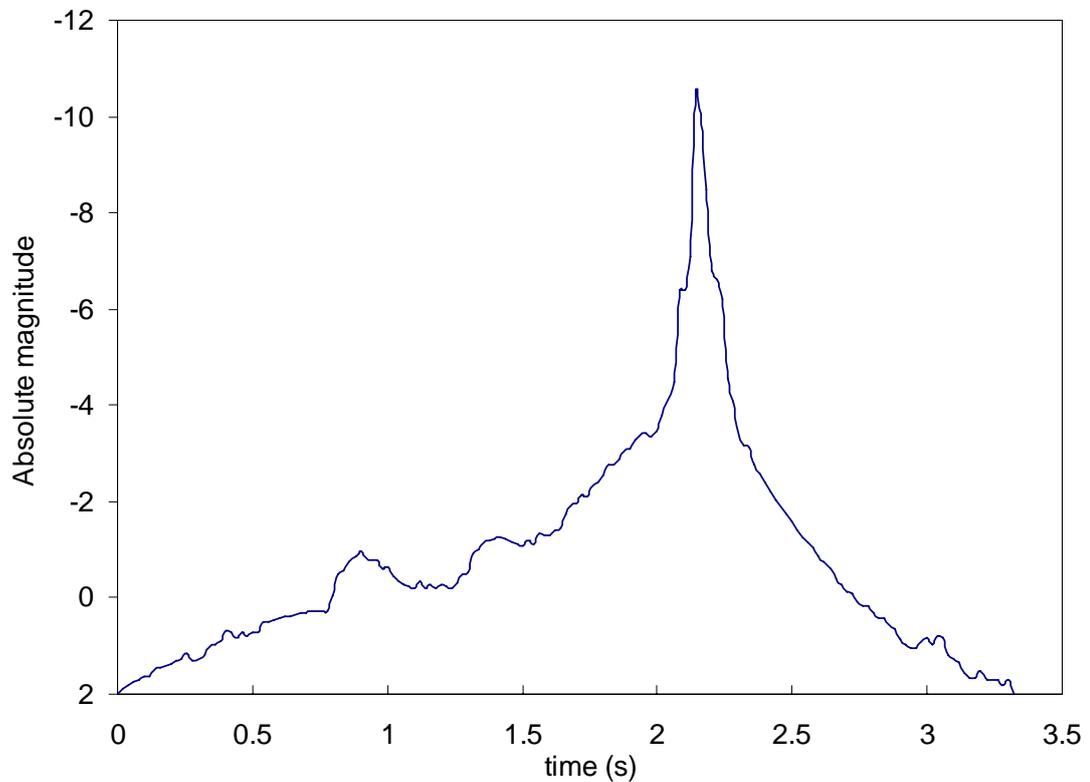

Figure 4. Lightcurve of the fireball analyzed in the text.

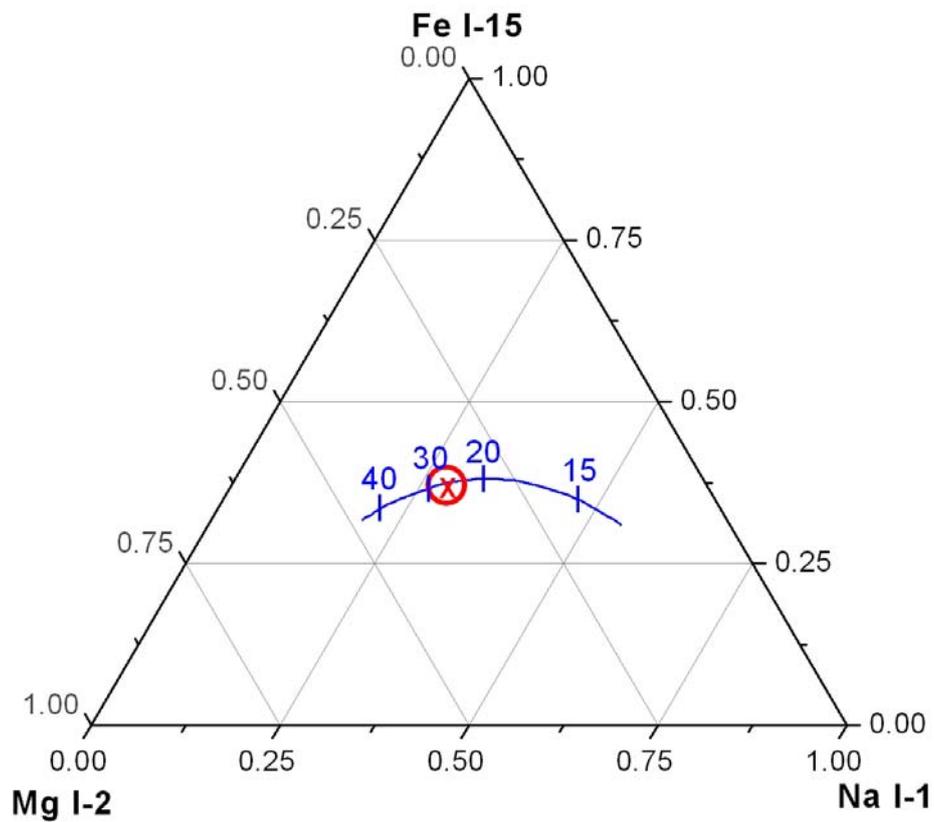

Figure 5. Solid line: expected relative intensity as a function of meteor velocity in km s$^{-1}$ of the Mg I-2, Na I-1 and Fe I-15 multiplets for chondritic meteoroids (Borovička et al. (2005)). Cross: experimental relative intensity calculated from the fireball spectrum Area inside the circle: uncertainty (error bars) of the calculated intensity.

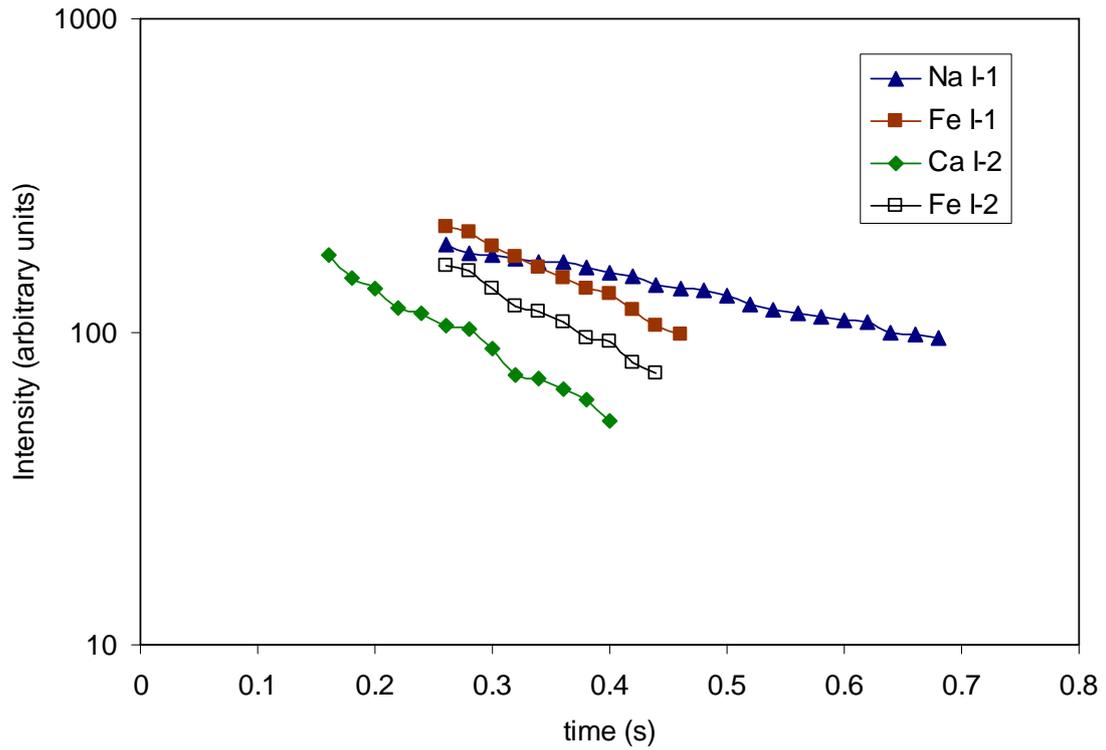

Figure 6. Evolution with time of the relative brightness of the main emission lines identified in the afterglow spectrum.

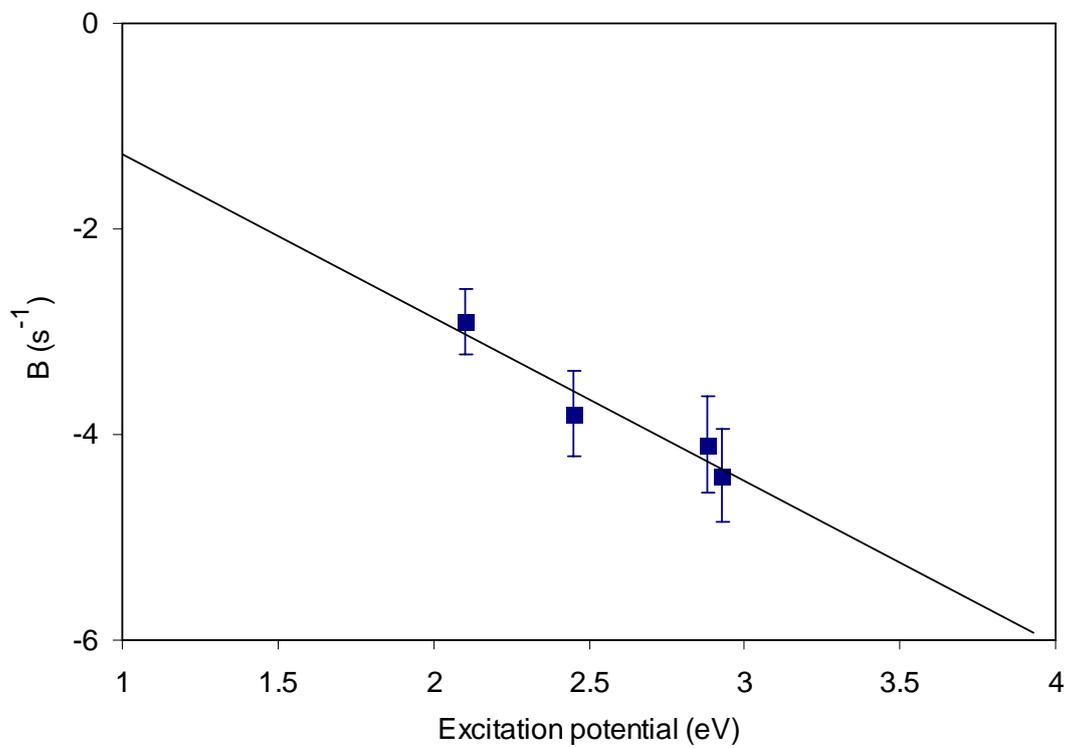

Figure 7. Dependence of the decay exponent B in Eq. (4) on the excitation potential. The solid line corresponds to the fit given by Eq. (5).